\pdfoutput=1
\PassOptionsToPackage{unicode}{hyperref}
\PassOptionsToPackage{hyphens}{url}
\PassOptionsToPackage{dvipsnames,svgnames,x11names}{xcolor}
\documentclass[
  11pt,
  letterpaper]{article}
\usepackage{xcolor}
\usepackage[margin=1in]{geometry}
\usepackage{amsmath,amssymb}
  \usepackage[T1]{fontenc}
  \usepackage[utf8]{inputenc}
  \usepackage{textcomp} 
\usepackage[]{newtxtext,newtxmath}
\IfFileExists{upquote.sty}{\usepackage{upquote}}{}
\IfFileExists{microtype.sty}{
  \usepackage[]{microtype}
  \UseMicrotypeSet[protrusion]{basicmath} 
}{}
\usepackage{setspace}
\makeatletter
\@ifundefined{KOMAClassName}{
  \IfFileExists{parskip.sty}{%
    \usepackage{parskip}
  }{
    \setlength{\parindent}{0pt}
    \setlength{\parskip}{6pt plus 2pt minus 1pt}}
}{
  \KOMAoptions{parskip=half}}
\makeatother
\usepackage{longtable,booktabs,array}
\usepackage{caption}
\usepackage{calc} 
\usepackage{etoolbox}
\makeatletter
\patchcmd\longtable{\par}{\if@noskipsec\mbox{}\fi\par}{}{}
\makeatother
\usepackage{graphicx}
\makeatletter
\newsavebox\pandoc@box
\newcommand*\pandocbounded[1]{
  \sbox\pandoc@box{#1}%
  \Gscale@div\@tempa{\textheight}{\dimexpr\ht\pandoc@box+\dp\pandoc@box\relax}%
  \Gscale@div\@tempb{\linewidth}{\wd\pandoc@box}%
  \ifdim\@tempb\p@<\@tempa\p@\let\@tempa\@tempb\fi
  \ifdim\@tempa\p@<\p@\scalebox{\@tempa}{\usebox\pandoc@box}%
  \else\usebox{\pandoc@box}%
  \fi%
}
\def\fps@figure{htbp}
\makeatother
\providecommand{\tightlist}{%
  \setlength{\itemsep}{0pt}\setlength{\parskip}{0pt}}
\usepackage{newunicodechar}
\newunicodechar{−}{\ensuremath{-}}
\newunicodechar{≈}{\ensuremath{\approx}}
\newunicodechar{≥}{\ensuremath{\geq}}
\newunicodechar{≤}{\ensuremath{\leq}}
\newunicodechar{→}{\ensuremath{\rightarrow}}
\newunicodechar{±}{\ensuremath{\pm}}
\newunicodechar{×}{\ensuremath{\times}}
\newunicodechar{Δ}{\ensuremath{\Delta}}
\newunicodechar{—}{---}
\newunicodechar{–}{--}
\newunicodechar{′}{'}
\usepackage{etoolbox}
\AtBeginEnvironment{longtable}{\scriptsize}
\renewenvironment{abstract}{\vspace{4pt}\begin{center}\large\bfseries Abstract\end{center}\vspace{-2pt}\normalsize\setlength{\parindent}{0pt}\setlength{\parskip}{6pt}}{\vspace{8pt}}
\usepackage{microtype}
\usepackage{sectsty}
\sectionfont{\large}
\subsectionfont{\normalsize}
\usepackage{bookmark}
\IfFileExists{xurl.sty}{\usepackage{xurl}}{} 
\makeatletter
\@ifundefined{xmpquote}{}{}
\makeatother
\hypersetup{
  pdftitle={Small Average Error{,} Wide Individual Error: General-Purpose LLMs Estimating Energy Intake from Meal Photos in a Feeding Trial},
  colorlinks=true,
  linkcolor={black},
  filecolor={Maroon},
  citecolor={Blue},
  urlcolor={blue},
  pdfcreator={LaTeX via pandoc}}

\title{Small Average Error, Wide Individual Error: General-Purpose LLMs Estimating Energy Intake from Meal
Photos in a Feeding Trial}
\author{Xiaomeng Li\\ {\normalsize Link Heart Limited, Houston, TX, USA}}
\date{Preprint, October 2026}

\begin{document}
\maketitle
\begin{abstract}
Photo-based estimation of food energy could reduce the burden of dietary assessment, and general-purpose
multimodal large language models (LLMs) make such estimates easy to obtain. Most evaluations use one photo
taken before eating and compare with the energy served, while dietary assessment needs the energy actually
eaten. We evaluated eight general-purpose models from three vendors (Claude Haiku, Sonnet and Opus 5.5; GPT-6
luna, GPT-6.1 sol and GPT-6 astra; Gemini 3.8 Flash and Gemini 4 Argon) on 802 meals from ACETADA, a
controlled feeding trial with weighed intake, giving each model the before photo only, before and after photos
in separate calls, or both photos in one request. Three questions and two limits were set in advance; the
further analyses are exploratory. Several results went against what we expected. Estimating the two photos
separately and subtracting under-estimated intake for six of eight models: the models still report hundreds of
kilocalories for a nearly empty tray. Much of the gain from the after photo could be had without it, by
scaling the before-only estimate by the average share eaten in other participants (0.87); beyond that
constant, the second photo lowered meal-level error by 2.2 to 4.4 points, mainly in meals with food left. With
one photo, two models showed a near-zero intake bias, but only because they under-estimated the served energy
and ignored the uneaten food. In ablations, adding the after photo improved accuracy under both prompt formats
for seven models, while prompt effects varied across models. Summed over 148 feeding days, five models kept
the 90\% confidence interval of their mean error within a pre-set ±10\% margin, and one (Gemini 3.8 Flash)
also reached the pre-set ±30\% limits of agreement, narrowly (−28\% to +27\%; the confidence intervals of both
limits cross 30\%, so limits outside ±30\% cannot be excluded); the other seven did not. Protein was
under-estimated by every model, most in protein-dense meals; repeated calls changed error by at most 1.4
points, and a higher reasoning effort helped only the smallest model (Haiku 5.5, 4.2 points) and changed the
others by less than 1 point. A low average bias is not evidence of accurate individual intake estimation, and
estimates of this kind should support, not replace, assessment by a dietitian.
\end{abstract}

\setstretch{1.05}
\section{Introduction}\label{introduction}

Diet records are used in the care of obesity and diabetes and in nutrition research, but self-reported intake
has well known errors and is a burden to collect {[}1{]}. Image-based methods were proposed to reduce this
burden {[}2{]}. General-purpose multimodal large language models (LLMs) can now return an energy estimate from
a meal photo without any task-specific training, and several studies have measured how well they do
{[}5,~12--14{]}.

Most of these evaluations give the model one photo taken before eating and compare its answer with the energy
served. Health questions, however, depend on the energy that was actually eaten. Work that uses photos taken
before and after eating either trains a dedicated model and reports grams {[}6{]}, or asks a general model
about each photo separately and subtracts, which was found to be unreliable {[}7{]}. LLM estimates have also
been validated against weighed food records, for a customized GPT-4o system used by dietetic trainees who
recorded their own meals {[}9{]}. To our knowledge, no study has given several general-purpose models both
photos in one request and asked for consumed energy and macronutrients. None has evaluated such estimates in a
controlled feeding trial at the level of a feeding day, where the same trial has published results for
established 24-hour recall methods {[}3{]}. That is what we do here.

We use the public ACETADA data from such a trial {[}4,~5{]} and ask three pre-registered questions:

\begin{itemize}
\tightlist
\item
  \textbf{Q1.} Does the after photo improve estimates of consumed energy compared with the before photo alone?
\item
  \textbf{Q2.} Is the change consistent across models?
\item
  \textbf{Q3.} Summed over a feeding day, are the mean error and the individual error small enough, judged
  against limits we set in advance?
\end{itemize}

We make three contributions. First, a paired comparison of three input conditions (before photo only; separate
calls with subtraction; both photos in one request) for eight general-purpose models from three vendors, on
802 meals with weighed consumed energy. It shows that subtraction fails and that a fixed-fraction baseline
matches much of the two-photo gain. Second, ablations that separate the second photo from the prompt (the
photo helped under both prompt formats; prompt effects varied by model), and a decomposition showing that a
small before-only intake bias can coexist with a large under-estimate of served energy; mean error alone is
therefore a misleading criterion. Third, day-level agreement with Bland--Altman limits and an
equivalence-style check of the mean error, together with macronutrient, repeatability and reasoning-effort
results. The trial's published recall results are given as a descriptive reference only.

\section{Related Work}\label{related-work}

\textbf{LLM estimates of food energy from photos.} Zero-shot multimodal LLMs under-estimate more as portions
grow {[}13{]} and keep high calorie error even when the foods are identified correctly {[}12{]}. On ACETADA,
the dataset paper evaluated several large multimodal models with the before photo and contextual metadata,
against consumed amounts {[}5{]}. Agentic decomposition into database-grounded food items has also been tried
on this data {[}14{]}.

\textbf{Before-and-after photos and leftovers.} Estimating leftovers from paired photos has a long history in
computer vision, for example in hospital settings with RGB-D image pairs {[}10{]} and in a university canteen
{[}20{]}; SNAPMe collected before and after photos with ASA24 records as the reference, not weighed intake
{[}11{]}. DietDelta trains a dedicated vision-language model on ACETADA photo pairs to predict the grams
eaten, with 14.2\% error relative to mean weight {[}6{]}. In the same paper, a closed general model given both
photos and the true list of foods had 38.9\% error, and estimating the two photos separately and subtracting
gave about the same. Nakagawa and Yamamoto studied energy estimates of 26 vision-language models and, on
ACETADA, estimated before and after photos in separate calls with the same prompt {[}7{]} (preprint). For
trays with at most 5\% of energy left, the after-photo estimate was still a median 39--83\% of the
before-photo estimate, depending on the model. They concluded that intake should not be derived by
subtraction. An early study used GPT-4V with before and after images from wearable cameras in a small
semi-free-living sample {[}8{]}.

\textbf{Validation against weighed intake.} A customized GPT-4o system was validated against three-day weighed
food records of 57 dietetic trainees (714 images), with Bland--Altman analysis and equivalence testing
{[}9{]}. The ACETADA trial itself compared four technology-assisted 24-hour recall methods with weighed intake
{[}3{]}.

\textbf{Our position.} We do not propose a new model. We evaluate general-purpose models in three input
conditions, on consumed energy and macronutrients, at the level of the feeding day in a controlled trial, and
we report the trial's recall results only as a descriptive reference.

\section{Data}\label{data}

\textbf{Source.} ACETADA comes from a randomized crossover controlled feeding study in Perth, Western
Australia {[}3,~4{]}. Adults ate laboratory-prepared breakfast, lunch and dinner on up to three feeding days.
Every food and drink was weighed before and after the meal, and nutrient values were assigned by dietitians.
The public release contains, for each meal, a photo before and a photo after eating, the weight of each item
before and after, the amount eaten, and energy, protein, carbohydrate and fat of the amount eaten {[}5{]}.
Photos were taken by hand in the study room, on a tray with a reference colour card, at varying angles.

\textbf{Cleaning.} Of 806 released meals we removed 2 rows that repeat the photo pair of another meal of the
same participant on the same day, 1 meal under 30 kcal (an inclusion threshold set in the pre-registration,
carried over from our earlier work on other datasets) and 1 meal whose before photo shows no food, leaving 802
meals from 152 participants. In 63 meals an item has a negative after weight (a weighing or tare error), so
the amount eaten is slightly larger than the amount served; together this is 0.17\% of all energy eaten, and
we keep these meals.

\textbf{Leftover.} We define the energy left as energy served minus energy eaten, for the whole meal. Energy
served per item is the energy eaten divided by the grams eaten, times the grams before eating. This is
undefined when an item was not eaten at all (energy per gram unknown); these 38 meals are left out of
leftover-based analyses only. Of the remaining 764 meals, the median share of energy left is 5.9\%
(interquartile range 1.7--16.5\%); 292 meals have at least 10\% left, the cut point fixed in the
pre-registration. We use energy rather than weight because leaving half of the vegetables and leaving half of
the meat mean different things for intake.

\textbf{Feeding days.} A complete feeding day has all three meals in the cleaned set: 148 days from 101
participants.

Table 1. Data used in this study.

{\def\LTcaptype{none} 
\begin{longtable}[]{@{}ll@{}}
\toprule\noalign{}
& Value \\
\midrule\noalign{}
\endhead
\bottomrule\noalign{}
\endlastfoot
Meals (participants) & 802 (152) \\
Energy eaten per meal, median (IQR), kcal & 622 (463--857) \\
--- breakfast / lunch / dinner (n) & 524 (286) / 623 (257) / 763 (259) \\
Meals with energy served defined & 764 \\
Share of energy left per meal, median (IQR) & 5.9\% (1.7--16.5\%) \\
Meals with ≥10\% of energy left & 292 \\
Complete feeding days (participants) & 148 (101) \\
Energy eaten per day, median (IQR), kcal & 2,017 (1,593--2,636) \\
\end{longtable}
}

\section{Methods}\label{methods}

\textbf{Models.} Claude Haiku 5.5, Sonnet 5.5 and Opus 5.5 (Anthropic API) and GPT-6 luna, GPT-6.1 sol and
GPT-6 astra (OpenAI API), called on 16 October 2026, and Gemini 3.8 Flash and Gemini 4 Argon (Google Gemini
API), added under a pre-registered rule and called on 20 October 2026 (Appendix B gives the order of the
runs). All main runs used low reasoning effort. We report results by model name. The vendors' models differ in
more than size, so we do not attribute differences to the number of parameters.

\textbf{Input conditions.} (1) \emph{Before only:} the before photo with a direct prompt for the energy of
everything on the plate; this assumes that everything was eaten. (2) \emph{Subtraction:} the same prompt on
the before photo and on the after photo in separate calls; eaten = before − after, floored at zero. This
follows {[}7{]}. (3) \emph{Both photos:} the before and after photos in one request, with a prompt that asks
for the energy served, the energy left over and the energy eaten, plus protein, carbohydrate and fat eaten.
The prompt does not list the foods. Condition (3) changes both the visual input and the output instructions
relative to condition (1). To separate the two, we added after the main results (exploratory) two conditions:
(3a) both photos with the simple prompt of condition (1), which also returns grams eaten; and (3b) the before
photo only with the structured prompt of condition (3). The photo effect is then (3) − (3b) with the
structured prompt held, and (3a) − (1) with the simple prompt held; the prompt effect is (3b) − (1) and (3) −
(3a). Photos are converted to PNG with the long side at most 1,024 pixels, which also removes embedded
metadata. GPT images are sent with detail ``high''; the Anthropic and Gemini APIs have no equivalent setting
and process the image at its given size, so the vendors may see the images at different effective resolutions.
We did not try to equalize this. Full prompts are in Appendix A; prompts are identical across models.

\textbf{Metrics.} At meal level: mean absolute error as a percentage of mean true energy (MAE\%) and bias
(total estimate minus total truth, as a percentage of total truth). For the share eaten, we compare the
model's own ratio of eaten to served with the true ratio (mean difference, mean absolute difference and
Spearman correlation; the correlation only shows ranking). At day level, for each participant's feeding day we
compute e\_day = 100 × (estimated − true energy eaten) / true energy eaten, summing the three meals. We report
the mean of e\_day with its 95\% confidence interval, 95\% Bland--Altman limits of agreement (LoA) {[}15{]},
and the share of days within 800 kJ (191 kcal) of truth. Because participants contribute one to three days, we
compute the LoA with the repeated-measures method of Bland and Altman {[}21{]} (one-way analysis of variance,
between- and within-participant components) and report the within-participant intraclass correlation of
e\_day; we also report LoA on the log scale as ratio limits. We plot e\_day against the true intake rather
than the mean of the two {[}22{]}. The LoA describe the spread of single feeding days; they do not describe
usual intake or the reliability of averages over several days.

\textbf{Thresholds.} Before any dual-photo call we set two limits for Q3: an absolute mean day error of at
most 10\%, and 95\% LoA within ±30\%. These are limits chosen by us for this study. The ±30\% limit is close
to what the best recall method reached in the same trial (an image-based record analysed by trained staff)
{[}3{]}. They are not accepted clinical standards. For the mean error we use an equivalence-style rule: the
90\% confidence interval must lie within ±10\% {[}16{]}.

\textbf{Statistics.} Confidence intervals come from 2,000 bootstrap resamples of participants. Differences
between input conditions are paired. To test whether the gain depends on leftovers, we compute the gain
separately in meals with ≥10\% and \textless10\% of energy left and bootstrap the difference between the two
gains. For Q2 we describe heterogeneity between models by the range and standard deviation of the eight
models' gains, with bootstrap intervals, and list all 28 pairwise differences (unadjusted) in Appendix F.
Point estimates are computed on the full sample; the bootstrap is used only for intervals. Some participants
contributed up to three feeding days; resampling participants keeps their days together. Confidence intervals
are not adjusted for multiple comparisons, so we emphasize effect sizes and intervals.

\textbf{Reliability and effort (exploratory).} For the 100 pilot meals we repeated condition (3) twice more
with fresh calls (three estimates per meal) and report the per-meal coefficient of variation and the range of
MAE\% across repeats; and we ran condition (3) once with high reasoning effort on the same meals.

\textbf{Fixed-fraction baseline (exploratory).} To check whether the both-photos protocol does more than lower
all estimates by a constant, we scale each before-only estimate by c = (energy eaten) / (energy served),
summed over the meals of all \emph{other} participants (leave-one-participant-out; c = 0.869--0.876). The rule
was fixed before computing.

\textbf{Failures.} Network errors are retried by the client; a syntactically invalid answer is not re-asked. A
rule fixed in advance scores such an unparsable answer as that system's worst relative error: w = max
\textbar ŷ − y\textbar{} / y over the system's parsed meals, and the failed meal gets the synthetic estimate y
× (1 + w), which is then used in day sums like any other meal. In the subtraction condition, a failure of
either call counts as a failure. Macronutrient fields have no synthetic value; days with a missing field are
left out of that nutrient's day analysis. Failures were at most 11 of 802 meals per system (Appendix A); no
meal failed for more than one system, so failures are not tied to particular photos. Because the rule
penalizes the system with the most failures (Sonnet 5.5, 11 meals), Table 2 also gives results on the subset
of meals with valid output in all three conditions.

\textbf{Pre-registration and deviations.} The questions Q1--Q3, prompts, the two limits, the 10\% leftover cut
point and a pilot rule were committed with a timestamp before any dual-photo call. The specific tests that
answer Q2 and the equivalence-style check for Q3 were chosen later. A 100-meal pilot with two models met the
pre-set continue rule, and the prompt was not changed, so pilot meals stay in. After a literature check, and
before the full run, we added the subtraction condition. The leftover-group difference test, the error
decomposition, the model comparison against Haiku, the equivalence check, the fixed-fraction baseline, the
kcal-scale and pilot-exclusion sensitivity analyses, the local recalibration and the breakdown by meal were
added after the main results were known, following review, and are exploratory. Because the full run depended
on the pilot meeting its rule, we also report results without the pilot meals (Appendix F). The two Gemini
models were added after the first version of this paper, under a selection rule and budget registered before
any Gemini call (Appendix B); they were run through every condition, including the exploratory ones, with the
same prompts and scripts.

\section{Results}\label{results}

\subsection{Q1: the after photo}\label{q1-the-after-photo}

Table 2 shows meal-level results. Against the before-only condition, which assumes that everything was eaten,
the both-photos protocol lowered MAE\% for six of eight models, by 2.8 to 9.6 points (Opus 5.5, GPT-6 luna,
GPT-6.1 sol, GPT-6 astra, Gemini 3.8 Flash, Gemini 4 Argon); the change was not clear for Sonnet 5.5 and
absent for Haiku 5.5. This comparison, however, favours the protocol by construction in meals with food left:
the before-only error contains the uneaten energy by definition. The more informative baseline is the
before-only estimate scaled by a fixed share eaten taken from other participants (c = 0.87). Against that
baseline the protocol was better by 2.2 to 4.4 points, with intervals excluding zero for seven models (Opus
5.5: −2.2, CI −4.8 to +0.3). In meals with ≥10\% of energy left, its gain over the constant was clear for
GPT-6 astra (−8.8 points), Gemini 4 Argon (−7.7), Opus 5.5 (−7.5), GPT-6.1 sol (−6.6) and Gemini 3.8 Flash
(−5.4), but not for Haiku 5.5, Sonnet 5.5 or GPT-6 luna; in mostly eaten meals there was no gain for any model
(Appendix F). On the subset of meals with valid output in all conditions, Sonnet 5.5's both-photos error is
19.3\% rather than 21.8\%, because its 11 failures are scored at its worst error; the other models change by
at most 0.6 points.

Table 2. Meal-level MAE\% (bias \%) by input condition, with the fixed-fraction baseline (before only × c, c =
share eaten in other participants). Δ = both photos − baseline, MAE\% points {[}95\% CI{]}, on all 802 meals
and in meals with ≥10\% of energy left (n = 292). Valid subset: meals with a parsable answer in all three
conditions for that model.

{\def\LTcaptype{none} 
\begin{longtable}[]{@{}
  >{\raggedright\arraybackslash}p{(\linewidth - 16\tabcolsep) * \real{0.1111}}
  >{\raggedright\arraybackslash}p{(\linewidth - 16\tabcolsep) * \real{0.1111}}
  >{\raggedright\arraybackslash}p{(\linewidth - 16\tabcolsep) * \real{0.1111}}
  >{\raggedright\arraybackslash}p{(\linewidth - 16\tabcolsep) * \real{0.1111}}
  >{\raggedright\arraybackslash}p{(\linewidth - 16\tabcolsep) * \real{0.1111}}
  >{\raggedright\arraybackslash}p{(\linewidth - 16\tabcolsep) * \real{0.1111}}
  >{\raggedright\arraybackslash}p{(\linewidth - 16\tabcolsep) * \real{0.1111}}
  >{\raggedright\arraybackslash}p{(\linewidth - 16\tabcolsep) * \real{0.1111}}
  >{\raggedright\arraybackslash}p{(\linewidth - 16\tabcolsep) * \real{0.1111}}@{}}
\toprule\noalign{}
\begin{minipage}[b]{\linewidth}\raggedright
Model
\end{minipage} & \begin{minipage}[b]{\linewidth}\raggedright
Before only
\end{minipage} & \begin{minipage}[b]{\linewidth}\raggedright
Before × c
\end{minipage} & \begin{minipage}[b]{\linewidth}\raggedright
Subtraction
\end{minipage} & \begin{minipage}[b]{\linewidth}\raggedright
Both photos
\end{minipage} & \begin{minipage}[b]{\linewidth}\raggedright
Both photos, valid subset (n)
\end{minipage} & \begin{minipage}[b]{\linewidth}\raggedright
Δ vs before only
\end{minipage} & \begin{minipage}[b]{\linewidth}\raggedright
Δ vs before × c
\end{minipage} & \begin{minipage}[b]{\linewidth}\raggedright
Δ vs before × c, ≥10\% left
\end{minipage} \\
\midrule\noalign{}
\endhead
\bottomrule\noalign{}
\endlastfoot
Haiku 5.5 & 24.7 (−6.7) & 27.3 & 36.5 (−33.5) & 24.9 (−14.0) & 24.1 (800) & +0.1 {[}−1.9,~+2.3{]} & −2.5
{[}−4.2,~−0.6{]} & −1.8 {[}−4.4,~+0.6{]} \\
Sonnet 5.5 & 24.2 (+0.9) & 24.9 & 25.1 (−18.4) & 21.8 (−2.9) & 19.3 (789) & −2.4 {[}−5.3,~+0.2{]} & −3.0
{[}−5.5,~−0.8{]} & −1.3 {[}−5.9,~+2.7{]} \\
Opus 5.5 & 30.1 (+21.7) & 22.7 & 21.2 (+1.8) & 20.5 (+9.0) & 20.5 (801) & −9.6 {[}−13.2,~−6.5{]} & −2.2
{[}−4.8,~+0.3{]} & −7.5 {[}−12.3,~−3.3{]} \\
GPT-6 luna & 23.9 (+4.3) & 23.7 & 26.2 (−17.6) & 21.0 (−6.1) & 21.0 (802) & −2.8 {[}−5.0,~−1.0{]} & −2.6
{[}−4.2,~−1.1{]} & −1.3 {[}−5.1,~+1.9{]} \\
GPT-6.1 sol & 26.9 (+14.8) & 22.3 & 23.7 (−10.7) & 19.5 (+1.2) & 19.1 (801) & −7.4 {[}−10.3,~−4.6{]} & −2.8
{[}−4.8,~−0.7{]} & −6.6 {[}−10.6,~−3.1{]} \\
GPT-6 astra & 27.6 (+16.4) & 22.4 & 22.5 (−7.7) & 19.0 (+5.3) & 19.0 (802) & −8.6 {[}−11.3,~−6.1{]} & −3.4
{[}−5.3,~−1.5{]} & −8.8 {[}−12.7,~−5.1{]} \\
Gemini 3.8 Flash & 21.9 (+7.9) & 20.4 & 18.5 (−8.1) & 16.7 (−2.2) & 16.7 (802) & −5.2 {[}−7.5,~−3.1{]} & −3.7
{[}−5.2,~−2.3{]} & −5.4 {[}−8.4,~−2.8{]} \\
Gemini 4 Argon & 27.9 (+17.4) & 22.6 & 21.2 (+2.7) & 18.2 (+3.0) & 18.2 (802) & −9.6 {[}−12.3,~−7.2{]} & −4.4
{[}−6.2,~−2.6{]} & −7.7 {[}−11.2,~−4.4{]} \\
\end{longtable}
}

\textbf{Photo or prompt?} Table 3 separates the two changes made in condition (3). With the structured prompt
held fixed, adding the after photo lowered MAE\% for seven of eight models (Sonnet 5.5 −5.1, Opus 5.5 −2.5,
GPT-6 luna −2.5, GPT-6.1 sol −6.7, GPT-6 astra −8.1, Gemini 3.8 Flash −5.3, Gemini 4 Argon −5.3 points; Haiku
5.5 −1.3, CI −2.8 to +0.4). With the simple prompt held fixed, the second photo had the same or a larger
effect (−3.1 to −9.7 points for the same seven models; Haiku 5.5 +0.2). Prompt effects varied across models.
With the before photo only, the structured prompt changed MAE\% by −0.7 to +2.7 points for five models, but
improved Opus 5.5 by 7.1 points and Gemini 4 Argon by 4.4 points, because it makes these models subtract a
predicted leftover from their over-estimate. With both photos, the simple prompt was as good as or slightly
better than the structured one for every model (Sonnet 5.5: 18.9\% vs 21.8\%, of which the structured prompt's
11 failures account for most). Adding the after photo therefore improved accuracy under both prompt formats
for seven models, while the prompt's contribution depended on the model; the structured prompt is needed for
the macronutrient outputs. Day-level results for the simple-prompt protocol are similar to those of condition
(3) (Appendix F). From the simple two-photo prompt the models also returned grams eaten; against weighed
grams, the error relative to mean weight was 18.6\% (Gemini 4 Argon) to 37.9\% (Haiku 5.5), with every model
under-estimating grams by 12--37\%. This is in the range that a closed model reached on the same photo pairs
when given the true list of foods (38.9\%) and well above a model trained on this data (14.2\%) {[}6{]}.

Table 3. Photo versus prompt (exploratory): meal-level MAE\% (bias \%) for four combinations of input and
prompt, and paired differences {[}95\% CI{]}, 802 meals.

{\def\LTcaptype{none} 
\begin{longtable}[]{@{}
  >{\raggedright\arraybackslash}p{(\linewidth - 16\tabcolsep) * \real{0.1111}}
  >{\raggedright\arraybackslash}p{(\linewidth - 16\tabcolsep) * \real{0.1111}}
  >{\raggedright\arraybackslash}p{(\linewidth - 16\tabcolsep) * \real{0.1111}}
  >{\raggedright\arraybackslash}p{(\linewidth - 16\tabcolsep) * \real{0.1111}}
  >{\raggedright\arraybackslash}p{(\linewidth - 16\tabcolsep) * \real{0.1111}}
  >{\raggedright\arraybackslash}p{(\linewidth - 16\tabcolsep) * \real{0.1111}}
  >{\raggedright\arraybackslash}p{(\linewidth - 16\tabcolsep) * \real{0.1111}}
  >{\raggedright\arraybackslash}p{(\linewidth - 16\tabcolsep) * \real{0.1111}}
  >{\raggedright\arraybackslash}p{(\linewidth - 16\tabcolsep) * \real{0.1111}}@{}}
\toprule\noalign{}
\begin{minipage}[b]{\linewidth}\raggedright
Model
\end{minipage} & \begin{minipage}[b]{\linewidth}\raggedright
(1) Before, simple prompt
\end{minipage} & \begin{minipage}[b]{\linewidth}\raggedright
(3b) Before, structured prompt
\end{minipage} & \begin{minipage}[b]{\linewidth}\raggedright
(3a) Both photos, simple prompt
\end{minipage} & \begin{minipage}[b]{\linewidth}\raggedright
(3) Both photos, structured prompt
\end{minipage} & \begin{minipage}[b]{\linewidth}\raggedright
Photo effect, structured prompt (3) − (3b)
\end{minipage} & \begin{minipage}[b]{\linewidth}\raggedright
Photo effect, simple prompt (3a) − (1)
\end{minipage} & \begin{minipage}[b]{\linewidth}\raggedright
Prompt effect, before only (3b) − (1)
\end{minipage} & \begin{minipage}[b]{\linewidth}\raggedright
Prompt effect, both photos (3) − (3a)
\end{minipage} \\
\midrule\noalign{}
\endhead
\bottomrule\noalign{}
\endlastfoot
Haiku 5.5 & 24.7 (−6.7) & 26.2 (−17.8) & 25.0 (−14.3) & 24.9 (−14.0) & −1.3 {[}−2.8,~+0.4{]} & +0.2
{[}−1.9,~+2.2{]} & +1.5 {[}−0.3,~+3.2{]} & −0.1 {[}−1.6,~+1.4{]} \\
Sonnet 5.5 & 24.2 (+0.9) & 26.9 (−1.3) & 18.9 (−8.0) & 21.8 (−2.9) & −5.1 {[}−9.3,~−1.3{]} & −5.3
{[}−8.1,~−3.0{]} & +2.7 {[}−1.3,~+6.9{]} & +2.9 {[}+1.2,~+4.6{]} \\
Opus 5.5 & 30.1 (+21.7) & 22.9 (+9.2) & 20.3 (+6.3) & 20.5 (+9.0) & −2.5 {[}−4.4,~−0.6{]} & −9.7
{[}−13.4,~−6.5{]} & −7.1 {[}−9.8,~−4.9{]} & +0.1 {[}−0.7,~+1.1{]} \\
GPT-6 luna & 23.9 (+4.3) & 23.5 (+3.7) & 20.7 (−7.1) & 21.0 (−6.1) & −2.5 {[}−4.1,~−1.0{]} & −3.1
{[}−5.1,~−1.4{]} & −0.4 {[}−1.5,~+0.7{]} & +0.3 {[}−0.7,~+1.3{]} \\
GPT-6.1 sol & 26.9 (+14.8) & 26.2 (+17.0) & 18.7 (−2.2) & 19.5 (+1.2) & −6.7 {[}−9.6,~−3.8{]} & −8.2
{[}−11.1,~−5.6{]} & −0.7 {[}−1.5,~+0.2{]} & +0.8 {[}−0.2,~+2.1{]} \\
GPT-6 astra & 27.6 (+16.4) & 27.2 (+19.1) & 18.4 (+1.6) & 19.0 (+5.3) & −8.1 {[}−10.9,~−5.6{]} & −9.2
{[}−12.1,~−6.6{]} & −0.5 {[}−1.4,~+0.4{]} & +0.6 {[}+0.0,~+1.2{]} \\
Gemini 3.8 Flash & 21.9 (+7.9) & 21.9 (+9.6) & 16.8 (−4.3) & 16.7 (−2.2) & −5.3 {[}−7.8,~−3.2{]} & −5.1
{[}−7.4,~−2.9{]} & +0.1 {[}−0.7,~+0.9{]} & −0.2 {[}−0.9,~+0.6{]} \\
Gemini 4 Argon & 27.9 (+17.4) & 23.5 (−5.4) & 19.0 (−1.7) & 18.2 (+3.0) & −5.3 {[}−7.2,~−3.3{]} & −8.8
{[}−11.7,~−6.0{]} & −4.4 {[}−7.3,~−1.4{]} & −0.8 {[}−1.9,~+0.2{]} \\
\end{longtable}
}

The models read the share eaten with little bias. From their own numbers, the ratio of eaten to served
differed from the true ratio by −0.3 to +2.6 percentage points on average, with a mean absolute difference of
5.3 to 7.8 points and Spearman correlations of 0.63 to 0.77. Subtraction under-estimated intake for six of
eight models (bias −7.7\% to −33.5\%), in line with {[}7{]}: the estimate for the after photo stays too high.
Opus 5.5 (+1.8\%) and Gemini 4 Argon (+2.7\%) were the exceptions; both over-estimate the served energy with
one photo, which offsets the too-high after-photo estimate.

\textbf{Day level.} The both-photos protocol had narrower day-level limits of agreement than the constant
baseline for seven models (for example Sonnet 5.5: −67\% to +51\% vs −41\% to +40\%; Haiku 5.5 was the
exception, 77 vs 82 points wide) and more days within 800 kJ for all eight, while the constant sometimes gave
a smaller mean error (Opus 5.5: +9.2\% vs +12.0\%) (Appendix F). The constant estimates a meal's share eaten
with a mean absolute error of 11.2 points, compared with 5.3--7.8 points for the models' own shares.

\textbf{Robustness.} On the kcal scale, the per-meal change \textbar both − eaten\textbar{} − \textbar before
− eaten\textbar{} was more favourable in meals with ≥10\% left than in mostly eaten meals for every model
(differences −43 to −138 kcal per meal, all intervals excluding zero; Appendix F); as noted above, part of
this difference follows from the definition of the before-only condition. Excluding the 100 pilot meals, or
all pilot participants, did not change the direction of any main result (Appendix F). Our before-only errors
(152--205 kcal per meal, mean absolute percentage error 24--33\%) are in the range that the dataset paper
reported for closed models with a plain prompt on the same photos (166--211 kcal, 23--34\%) {[}5{]}.

\subsection{Q2: differences between models}\label{q2-differences-between-models}

The direction of the change was the same for seven of eight models and clear for six, but its size was not
consistent: across the eight models the gains against the before-only condition ranged over 9.8 points (95\%
CI 7.9 to 13.0; standard deviation 3.5 points, CI 2.7 to 4.4). Of the 28 pairwise differences, 21 had
intervals excluding zero (Appendix F); Opus 5.5, Gemini 4 Argon, GPT-6 astra and GPT-6.1 sol gained most,
Haiku 5.5, Sonnet 5.5 and GPT-6 luna least. Against the fixed-fraction baseline the gains were more alike (2.2
to 4.4 points), so much of the heterogeneity reflects how much each model over-estimated intake with the
before photo alone.

The before-only results help explain this (Table 4). The error of a before-only estimate against energy eaten
is the sum of two parts: the error against energy served, and the energy left. The true energy left cannot be
negative; in our labels it is slightly negative for a few meals because of weighing error (Section 3), and on
average it is +103 kcal per meal. Opus 5.5, GPT-6.1 sol, GPT-6 astra and Gemini 4 Argon estimated the served
energy with little bias, so the leftover appeared as an over-estimate of intake. Haiku 5.5, Sonnet 5.5, GPT-6
luna and Gemini 3.8 Flash under-estimated the served energy, and this under-estimate offset the leftover; for
Sonnet 5.5 the two parts almost cancelled (−104 and +103 kcal per meal). The offset also happened within the
same meal: among meals with ≥10\% left, Haiku 5.5 under-estimated the served energy and still ended closer to
the energy eaten in 178 of 292 meals, by a median of 137 kcal. These results show that a small before-only
intake bias can coexist with a substantial under-estimate of served energy, partly offset by uneaten food.
This decomposition does not isolate why predictions change under the both-photos protocol, which may also
change the model's estimate of the served energy.

Table 4. Decomposition of the before-only error, kcal per meal (764 meals with energy served defined).

{\def\LTcaptype{none} 
\begin{longtable}[]{@{}
  >{\raggedright\arraybackslash}p{(\linewidth - 10\tabcolsep) * \real{0.1667}}
  >{\raggedright\arraybackslash}p{(\linewidth - 10\tabcolsep) * \real{0.1667}}
  >{\raggedright\arraybackslash}p{(\linewidth - 10\tabcolsep) * \real{0.1667}}
  >{\raggedright\arraybackslash}p{(\linewidth - 10\tabcolsep) * \real{0.1667}}
  >{\raggedright\arraybackslash}p{(\linewidth - 10\tabcolsep) * \real{0.1667}}
  >{\raggedright\arraybackslash}p{(\linewidth - 10\tabcolsep) * \real{0.1667}}@{}}
\toprule\noalign{}
\begin{minipage}[b]{\linewidth}\raggedright
Model
\end{minipage} & \begin{minipage}[b]{\linewidth}\raggedright
Estimate − served
\end{minipage} & \begin{minipage}[b]{\linewidth}\raggedright
Energy left
\end{minipage} & \begin{minipage}[b]{\linewidth}\raggedright
Estimate − eaten
\end{minipage} & \begin{minipage}[b]{\linewidth}\raggedright
Meals with ≥10\% left where the two parts offset*
\end{minipage} & \begin{minipage}[b]{\linewidth}\raggedright
Median offset, kcal
\end{minipage} \\
\midrule\noalign{}
\endhead
\bottomrule\noalign{}
\endlastfoot
Haiku 5.5 & −156 & +103 & −53 & 178 / 292 & 153 \\
Sonnet 5.5 & −104 & +103 & −1 & 146 / 292 & 147 \\
Opus 5.5 & +38 & +103 & +141 & 61 / 292 & 126 \\
GPT-6 luna & −80 & +103 & +23 & 141 / 292 & 153 \\
GPT-6.1 sol & −8 & +103 & +95 & 84 / 292 & 141 \\
GPT-6 astra & +4 & +103 & +107 & 88 / 292 & 126 \\
Gemini 3.8 Flash & −55 & +103 & +48 & 111 / 292 & 140 \\
Gemini 4 Argon & +12 & +103 & +115 & 71 / 292 & 133 \\
\end{longtable}
}

*The served energy was under-estimated and the estimate was closer to the energy eaten than to the energy
served. Offset = \textbar estimate − served\textbar{} − \textbar estimate − eaten\textbar{} in those meals.

\subsection{Q3: agreement over a feeding day}\label{q3-agreement-over-a-feeding-day}

Table 5 and Figure 1 show day-level results. With both photos, the 90\% confidence interval of the mean day
error was within ±10\% for Sonnet 5.5 (−0.6\%), Gemini 3.8 Flash (−0.5\%), GPT-6.1 sol (+3.4\%), GPT-6 luna
(−4.3\%) and Gemini 4 Argon (+4.5\%); GPT-6 astra was just outside (+7.6\%, upper limit 10.3\%). Haiku 5.5
(−12.0\%) and Opus 5.5 (+12.0\%) were outside. One system met the ±30\% limits for individual agreement:
Gemini 3.8 Flash, with limits of −28\% to +27\%. It met them narrowly: the bootstrap 95\% confidence intervals
of the two limits were −32\% to −23\% and +20\% to +32\%, so limits outside ±30\% cannot be excluded, and the
same model with the simple two-photo prompt gave −28\% to +24\% (Appendix F). For the other seven systems the
limits spanned 59 to 82 percentage points; Gemini 4 Argon (−25\% to +34\%) and GPT-6 luna (−38\% to +30\%)
were closest to the pre-set limits, and GPT-6.1 sol (−31\% to +38\%) and GPT-6 astra (−27\% to +42\%) had
similar widths but were placed differently. Between 41\% and 57\% of days were within 800 kJ of truth for the
seven models whose meal-level error fell. Even for the model within the limits, a single feeding day can be
off by about a quarter of its energy in either direction. With the before photo only, Haiku 5.5 also passed
the equivalence check (−4.2\%, 90\% CI −8.1 to −0.3), in line with the pattern seen in the meal-level
decomposition (Section 5.2). A small mean error alone therefore does not show that a method reads intake
correctly.

Table 5. Day-level agreement for energy eaten (148 days, 101 participants). Mean = mean of day \% errors
{[}95\% CI{]}. Equivalence: 90\% CI within ±10\%.

{\def\LTcaptype{none} 
\begin{longtable}[]{@{}llllll@{}}
\toprule\noalign{}
Model, input & Mean day error \% & 90\% CI & Equivalence & 95\% LoA \% & Days within 800 kJ \% \\
\midrule\noalign{}
\endhead
\bottomrule\noalign{}
\endlastfoot
Haiku 5.5, before only & −4.2 {[}−8.9,~+0.7{]} & −8.1, −0.3 & yes & −48 to +39 & 34 \\
Haiku 5.5, both photos & −12.0 {[}−16.2,~−7.2{]} & −15.7, −8.2 & no & −53 to +29 & 29 \\
Sonnet 5.5, before only & +5.5 {[}−0.7,~+12.3{]} & +0.1, +11.0 & no & −62 to +73 & 37 \\
Sonnet 5.5, both photos & −0.6 {[}−4.2,~+3.2{]} & −3.7, +2.5 & yes & −41 to +40 & 41 \\
Opus 5.5, before only & +25.4 {[}+19.4,~+32.3{]} & +20.3, +31.0 & no & −32 to +83 & 25 \\
Opus 5.5, both photos & +12.0 {[}+8.7,~+15.5{]} & +9.2, +14.8 & no & −25 to +49 & 43 \\
GPT-6 luna, before only & +6.7 {[}+1.9,~+11.8{]} & +2.6, +10.9 & no & −41 to +54 & 39 \\
GPT-6 luna, both photos & −4.3 {[}−7.4,~−1.0{]} & −6.9, −1.5 & yes & −38 to +30 & 43 \\
GPT-6.1 sol, before only & +17.7 {[}+12.6,~+23.4{]} & +13.1, +22.5 & no & −33 to +69 & 29 \\
GPT-6.1 sol, both photos & +3.4 {[}+0.3,~+6.5{]} & +0.8, +6.0 & yes & −31 to +38 & 51 \\
GPT-6 astra, before only & +20.2 {[}+15.0,~+25.9{]} & +15.6, +25.0 & no & −31 to +71 & 24 \\
GPT-6 astra, both photos & +7.6 {[}+4.4,~+10.8{]} & +5.0, +10.3 & no & −27 to +42 & 45 \\
Gemini 3.8 Flash, before only & +10.6 {[}+5.8,~+16.0{]} & +6.4, +15.0 & no & −36 to +57 & 39 \\
Gemini 3.8 Flash, both photos & −0.5 {[}−2.9,~+2.0{]} & −2.5, +1.6 & yes & −28 to +27 & 57 \\
Gemini 4 Argon, before only & +20.4 {[}+14.9,~+26.4{]} & +15.8, +25.4 & no & −32 to +72 & 28 \\
Gemini 4 Argon, both photos & +4.5 {[}+1.9,~+7.1{]} & +2.3, +6.7 & yes & −25 to +34 & 55 \\
\end{longtable}
}

\begin{figure}
\centering
\includegraphics[width=1\linewidth,height=\textheight,keepaspectratio,alt={Day-level Bland--Altman plots for energy eaten,~both photos in one request. Each point is one participant's feeding day. Solid and dashed orange lines: mean error and 95\% limits of agreement. Grey lines: the pre-set ±10\% (solid) and ±30\% (dotted) limits. Black dashed line: local-linear smoother.}]{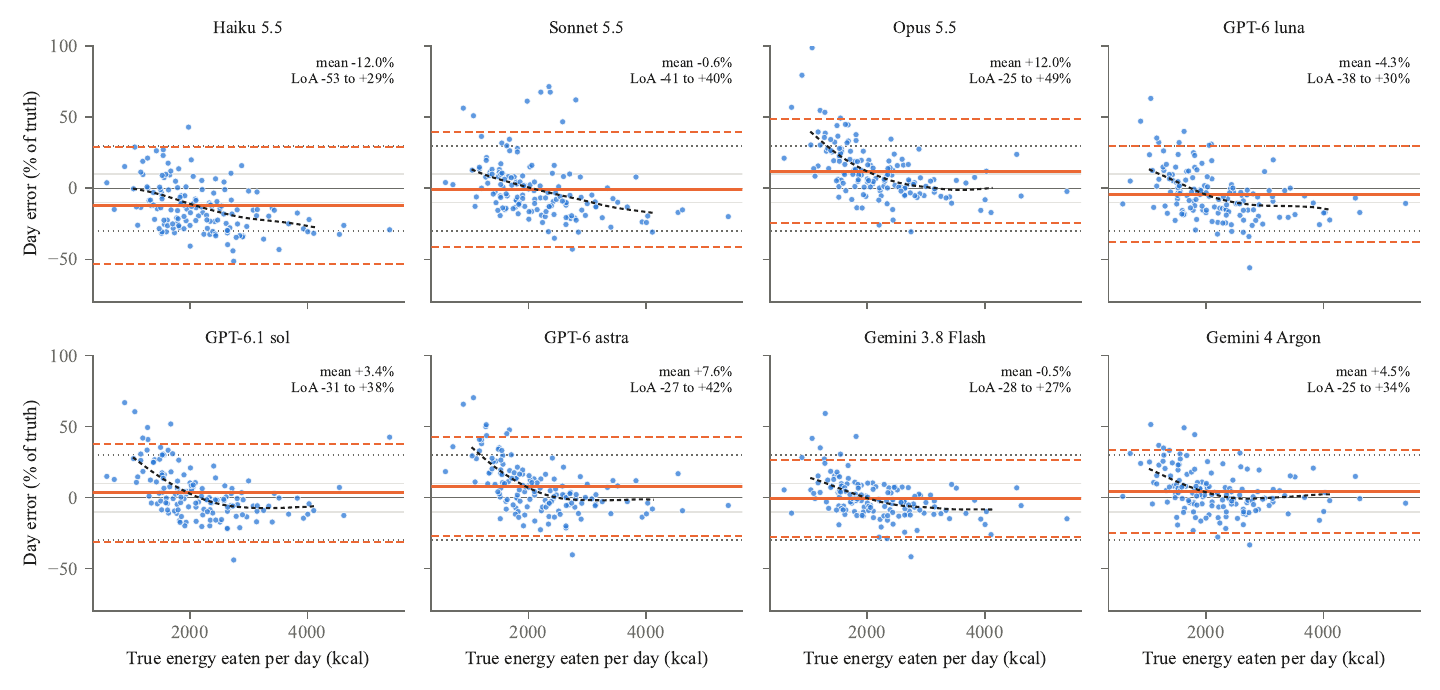}
\caption{Day-level Bland--Altman plots for energy eaten,~both photos in one request. Each point is one
participant's feeding day. Solid and dashed orange lines: mean error and 95\% limits of agreement. Grey lines:
the pre-set ±10\% (solid) and ±30\% (dotted) limits. Black dashed line: local-linear smoother.}
\end{figure}

Repeated days from the same participant were correlated (intraclass correlation of e\_day 0.15 to 0.53); the
repeated-measures limits {[}21{]} were the same as the naive ones to the nearest point, because the within-
and between-participant components add up to the total variance. On the log scale the limits are ratio limits
of about ×0.67 to ×1.4 for the better models and ×0.76 to ×1.29 for Gemini 3.8 Flash (Appendix F). In Figure 1
the day error falls as the true intake rises for every model (smoothed lines): days with low intake were
over-estimated and days with high intake under-estimated, by 11 to 22 percentage points per doubling of true
intake. On the log scale the slope of the estimate on the truth was 0.74 to 0.86 at day level and 0.68 to 0.84
at meal level; the two Gemini models shrank least. This is the pattern of shrinkage toward a typical meal
rather than a specific failure on large portions; it also means that percentage limits are not constant across
the intake range. Subtraction gave a small mean error for Opus 5.5 (+3.7\%, LoA −29\% to +37\%) and Gemini 4
Argon (+3.9\%, LoA −25\% to +33\%) but large negative errors for the other six models (Appendix C).

For orientation only, Table 6 lists the energy results that the trial reported for four 24-hour recall methods
{[}3{]}. They come from the same trial and the same reference (weighed intake), but from different feeding
days, possibly different participants, and a statistic computed by the trial's authors. The trial reports a
mean and standard deviation of the ratio of estimated to true intake; an approximate error range derived from
these rounded statistics is given in Appendix F for completeness. We do not compare methods.

Table 6. Published recall results from the same trial (descriptive reference; not compared with our results).

{\def\LTcaptype{none} 
\begin{longtable}[]{@{}lll@{}}
\toprule\noalign{}
Method {[}3{]} & Feeding days & Estimated / true energy, mean (SD) \% \\
\midrule\noalign{}
\endhead
\bottomrule\noalign{}
\endlastfoot
ASA24-Australia (self-administered) & 143 & 105 (29) \\
Intake24-Australia (self-administered) & 150 & 102 (29) \\
Mobile food record, trained analyst & 148 & 101 (15) \\
Image-assisted interviewer recall & 145 & 115 (20) \\
\end{longtable}
}

\subsection{Macronutrients (secondary)}\label{macronutrients-secondary}

Table 7. Day-level mean \% error (95\% LoA) for macronutrients eaten, both photos.

{\def\LTcaptype{none} 
\begin{longtable}[]{@{}llll@{}}
\toprule\noalign{}
Model & Protein & Carbohydrate & Fat \\
\midrule\noalign{}
\endhead
\bottomrule\noalign{}
\endlastfoot
Haiku 5.5 & −29 (−70 to +12) & +1 (−40 to +43) & −12 (−52 to +29) \\
Sonnet 5.5 & −24 (−62 to +14) & +14 (−33 to +62) & +3 (−39 to +45) \\
Opus 5.5 & −8 (−52 to +37) & +25 (−24 to +75) & +18 (−33 to +70) \\
GPT-6 luna & −16 (−63 to +30) & +12 (−31 to +55) & +0 (−45 to +46) \\
GPT-6.1 sol & −13 (−57 to +30) & +16 (−26 to +58) & +7 (−36 to +51) \\
GPT-6 astra & −10 (−54 to +35) & +21 (−23 to +64) & +12 (−31 to +56) \\
Gemini 3.8 Flash & −20 (−53 to +13) & +11 (−25 to +47) & +5 (−35 to +45) \\
Gemini 4 Argon & −3 (−46 to +40) & +11 (−29 to +51) & +10 (−34 to +54) \\
\end{longtable}
}

Protein was under-estimated by every model (−3\% to −29\% at day level; −7\% to −34\% of total protein at meal
level). The under-estimate was larger in protein-dense meals: across meals, the protein error correlated
negatively with protein density (Spearman −0.38 to −0.50 across models) and with the share of energy from
animal-source items (−0.09 to −0.35), and the bias was 6 to 17 percentage points larger in meals where
animal-source items supplied at least a third of the energy than in other meals (Appendix F). The models
appear to assign typical, rather than observed, protein contents. Carbohydrate errors were positive for seven
models (+11\% to +25\%) and near zero for Haiku 5.5 (+1\%); we did not test these differences. Limits of
agreement were wider than for energy. A model that estimates energy well does not necessarily estimate the
nutrients well.

\subsection{Exploratory: reliability and reasoning effort}\label{exploratory-reliability-and-reasoning-effort}

Hosted models are stochastic. On the 100 pilot meals, three fresh calls of condition (3) gave MAE\% values
that differed by at most 1.4 points within a model (Haiku 5.5 28.2--29.3; Sonnet 5.5 20.1--21.4; Opus 5.5
21.9--23.3; GPT-6 luna 22.4--23.7; GPT-6.1 sol 20.5--21.2; GPT-6 astra 20.5--21.1; Gemini 3.8 Flash
18.9--19.8; Gemini 4 Argon 18.8--20.2). The median per-meal coefficient of variation was 3.5--8.2\%, and two
calls on the same meal differed by 6--12\% of the true energy on average. Call-to-call noise is therefore
smaller than the photo effects in Table 3 for the seven models that improved, but of the same size as the
smallest differences we report (1--2 points). A high reasoning effort on the same meals helped one model:
Haiku 5.5 improved by 4.2 points (28.8\% to 24.6\%), while the other seven changed by less than 1 point (Opus
5.5 −0.4, Sonnet 5.5 −0.3, Gemini 3.8 Flash −0.1, Gemini 4 Argon 0.0, GPT-6.1 sol +0.3, GPT-6 astra +0.5,
GPT-6 luna +0.9), at 1.1 to 4.3 times the cost per meal (Appendix F). For the larger models, more reasoning
did not buy more accuracy on this task.

\subsection{Exploratory: does a small local sample help?}\label{exploratory-does-a-small-local-sample-help}

We refit two constants of a power law on the log scale (an intercept and a slope, as in the updating of
clinical prediction models {[}18,~19{]}) on 20 meals from other participants and applied it to the both-photos
estimates (200 random draws). The mean day error moved toward zero for most models (for example Opus 5.5 from
+12.1\% to +0.5\%, GPT-6 astra from +7.7\% to −1.1\%, Gemini 4 Argon from +4.5\% to +0.1\%). The width of the
limits of agreement changed in both directions (narrower for Opus 5.5 by 7 points and GPT-6 astra by 4 points;
wider for Haiku 5.5 by 9 points and GPT-6 luna by 4 points; about the same for the others) while their centre
moved toward zero; meal-level MAE\% went up slightly for seven models and down for Opus 5.5; and the share of
days within 800 kJ fell for some models (GPT-6.1 sol from 51\% to 41\%, Gemini 4 Argon from 56\% to 47\%).
Because calibration meals' participants are left out, these comparisons use a slightly smaller set of days
than Table 5 (hence, for example, +12.1\% instead of +12.0\% for Opus 5.5 without calibration). The metrics
moved in different directions, so in this trial a small local sample mainly corrected the mean (Appendix D).

\section{Discussion}\label{discussion}

\textbf{Main finding.} In this controlled feeding trial, the intuitive ways of using and judging a photo-based
estimator both misled. Estimating the two photos separately and subtracting failed, as previously reported
{[}7{]}, because the models do not read an almost empty tray as almost zero. Giving both photos in one request
did help, for seven of eight models, under both prompt formats, while prompt effects varied across models. But
a constant discount for the share eaten reproduced much of it. What the photo added on its own was 2 to 4
points of meal-level error, concentrated in meals with food left, and narrower day-level limits. One model,
Gemini 3.8 Flash, met both pre-set day-level limits, the ±30\% limits only narrowly; the vendor's flagship
model, Gemini 4 Argon, had the second-lowest meal-level error but limits of −25\% to +34\%. As with Sonnet 5.5
and Opus 5.5, the larger model of a vendor was not the more accurate one on this task. The clearest lesson is
about evaluation: a model can show a near-zero intake bias with one photo because two errors cancel, and a
method can pass a mean-error criterion while single-day errors remain wide.

\textbf{Scope.} The photos were taken under study instructions, by hand, on a tray with a reference card, and
the menu came from one trial kitchen. Free-living photos, other cuisines and other populations may give
different results. Meeting our pre-set limits for the mean in this sample does not mean the method meets the
needs of dietary surveys in general: we did not assess intake distributions, subgroup bias or usual intake.

\textbf{Mean error can mislead.} Small before-only intake biases came from models that under-estimated the
served energy, a pattern consistent with offset by uneaten food. The larger gain of the both-photos protocol
in meals with food left is partly a matter of definition, because the before-only condition assumes that
everything was eaten; the comparison with a constant discount is the one that shows whether a model read the
leftover. A study that tests one model with one input condition, and reports only the mean error, can draw the
wrong conclusion. We suggest reporting the limits of agreement, testing more than one model, and, where
possible, splitting the error against served energy from the leftover.

\textbf{Sources of the remaining error.} Our design cannot say how much of the remaining error comes from
portion size, food identification, energy density or label error. The fall of the percentage error with rising
intake, and log-scale slopes of about 0.75, point to shrinkage toward a typical meal, not to a specific
failure on large portions. The protein result, where protein-dense meals are under-estimated most, fits the
same reading.

\textbf{Practical meaning.} Two photos per meal are more work than one, so this approach fits research and
supervised settings better than everyday self-tracking. Because individual error is wide, even for the one
model within our pre-set limits, single-day estimates for one person should be used with care, and the protein
under-estimate matters for people who need a protein target. A systematic error in either direction could be
harmful for people with or at risk of eating disorders, and for people with low appetite on GLP-1 receptor
agonist drugs. Estimates of this kind should support, not replace, assessment by a dietitian.

\textbf{Relation to recall methods.} The recall results in Table 6 come from the same trial but not from the
same days, and they are summarized by a different team. We list them so that readers can see the range that
established methods reached under the same reference; this does not show that one approach is better.

\section{Limitations}\label{limitations}

\begin{itemize}
\tightlist
\item
  Recall data for the same participant-days are not public, so we could not compare methods directly.
\item
  Labels are weighed amounts with nutrient values assigned by dietitians; small weighing errors exist
  (negative after weights in 63 meals).
\item
  One trial, one city, one menu; photos under study instructions.
\item
  Hosted models change over time; we report model names and call dates, and release all outputs.
\item
  Several analyses (Section 4) were added after the main results and are exploratory; leftover-based analyses
  exclude 38 meals with undefined served energy.
\item
  We used one prompt per condition in the main runs; the ablations in Table 3 vary the prompt and the photos
  separately but test only two prompts, and other prompts or settings may do better. On 100 meals, a high
  reasoning effort improved only Haiku 5.5.
\item
  The Gemini models were added after the first version of this paper; their selection rule and budget were
  registered before any call, but the choice to add a third vendor was made with knowledge of the first six
  models' results, and the top-tier Gemini model was chosen by interpreting the registered rule (Appendix B).
\item
  Many tests are exploratory and unadjusted for multiple comparisons.
\item
  The full run was made conditional on a pilot result (Appendix B), so the existence of this study depends on
  the pilot having met its rule; the main results without the pilot meals are in Appendix F and are unchanged
  in direction.
\item
  Percentage errors shrink with intake, so the percentage limits of agreement are not constant across the
  intake range; log-scale limits are given in Appendix F.
\end{itemize}

\section{Conclusion}\label{conclusion}

In this controlled feeding trial, estimating intake by subtracting an after-photo estimate from a before-photo
estimate failed, and a constant consumption discount matched much of what a dual-photo protocol added; the
protocol still improved consumed-energy estimates for several general-purpose multimodal LLMs, particularly
for meals with substantial leftovers. Small mean daily errors coexisted with wide individual limits of
agreement (one of eight models met the pre-set ±30\% limits, and only narrowly), systematic protein
underestimation, and, for one photo, error cancellation that made an inaccurate estimator look accurate. These
findings show that improvement over an assumption of complete consumption, or low average bias alone, is
insufficient evidence of accurate individual intake estimation.

\section*{Data and Code Availability}\label{data-and-code-availability}
\addcontentsline{toc}{section}{Data and Code Availability}

\textbf{Data.} ACETADA {[}5{]} is public under CC BY-NC 4.0 for non-commercial research. We do not
redistribute images. We release scripts that download the data and rebuild our labels, the list of excluded
meals with reasons, and participant ids for the bootstrap.

\textbf{Code.} Code, prompts, the pre-registration file and per-meal outputs are available at
https://github.com/xtea/llm-intake-validation. It contains the prompts, the evaluation and analysis scripts,
the pre-registration file with its commit history, per-meal outputs of every model and condition, and the
scripts for all tables and figures. Because model responses are cached, all tables and figures can be rebuilt
without API calls. The code is released under the MIT license. The README lists the Python version (3.12),
pinned package versions (anthropic 0.84.0, openai 2.28.0, google-genai 2.31.0, matplotlib 3.11.2) and example
commands.

\section*{Institutional Review Board (IRB)}\label{institutional-review-board-irb}
\addcontentsline{toc}{section}{Institutional Review Board (IRB)}

The research does not require IRB approval. We analyze only a public, de-identified dataset released for
research; we had no interaction with any participant and received no identifiable private information, so the
work is not human subjects research under the U.S. Common Rule (45 CFR 46.102(e)). The original trial was
approved by the Curtin University Human Research Ethics Office (HRE2019-0222) and registered with the
Australian New Zealand Clinical Trials Registry (ACTRN12621000209897) {[}4{]}. Photos show a numbered study
card but no faces.

\section*{Acknowledgments and Disclosure}\label{acknowledgments-and-disclosure}
\addcontentsline{toc}{section}{Acknowledgments and Disclosure}

This work was supported by Link Heart Limited, which provided the research budget for the experiments. No
other funding was received. The author is co-founder and CTO of Link Heart Limited, which develops nutrition
software; the study uses only public data and was carried out as non-commercial research, independent of any
product. We thank the ACETADA investigators for releasing their data.

\section*{Statement on AI Use}\label{statement-on-ai-use}
\addcontentsline{toc}{section}{Statement on AI Use}

LLMs are the object of this study and were also used as tools.

\begin{enumerate}
\def\labelenumi{\arabic{enumi}.}
\tightlist
\item
  \textbf{Models under study.} Claude Haiku/Sonnet/Opus 5.5, GPT-6 luna/6.1 sol/6 astra and Gemini 3.8 Flash/4
  Argon produced all estimates (Section 4). \emph{Check:} metrics are computed by deterministic scripts from
  stored outputs.
\item
  \textbf{Code.} Evaluation and analysis scripts were written with help from Claude Opus 5.5 in Claude Code
  (version 2.1). \emph{Check:} the author reviewed the code; key numbers were recomputed by separate scripts
  and checked by hand against raw outputs.
\item
  \textbf{Literature search and references.} LLM assistance was used to find related work. \emph{Check:} every
  reference was matched to its publisher, DOI, PubMed, arXiv or medRxiv record, and each cited claim was
  checked against the source.
\item
  \textbf{Writing.} The author wrote the paper and is responsible for its content and claims; Claude Opus 5.5
  was used as a writing assistant for English drafting and editing (the author is not a native English
  speaker). \emph{Check:} the author reviewed and revised every sentence; all numbers come from the scripts
  above.
\end{enumerate}

\section*{Appendix A. Prompts and settings}\label{appendix-a.-prompts-and-settings}
\addcontentsline{toc}{section}{Appendix A. Prompts and settings}

\textbf{Before only and subtraction} (same text for each photo):

\begin{quote}
\texttt{This\ is\ a\ photo\ of\ a\ meal.\ Estimate\ the\ total\ food\ energy\ of\ everything\ on\ the\ plate.}\strut \\
\texttt{Reply\ with\ JSON\ only:\ \{"kcal":\ \textless{}number\textgreater{},\ "mass\_g":\ \textless{}total\ grams\ of\ food\textgreater{}\}}
\end{quote}

\textbf{Both photos} (before photo first, after photo second, one request):

\begin{quote}
\texttt{The\ first\ photo\ shows\ a\ meal\ before\ eating.\ The\ second\ photo\ shows\ the\ same\ tray\ after\ the\ person\ finished\ eating.}\strut \\
\texttt{Estimate\ the\ food\ energy\ that\ was\ served,\ the\ food\ energy\ that\ was\ left\ over,\ and\ the\ food\ energy\ the\ person\ actually\ ate}\strut \\
\texttt{(served\ minus\ left\ over).\ Also\ estimate\ the\ protein,\ carbohydrate\ and\ fat\ that\ the\ person\ actually\ ate.}\strut \\
\texttt{Reply\ with\ JSON\ only:\ \{"served\_kcal":\ \textless{}number\textgreater{},\ "leftover\_kcal":\ \textless{}number\textgreater{},\ "kcal":\ \textless{}number\ eaten\textgreater{},\ "protein\_g":\ \textless{}number\textgreater{},\ "carbs\_g":\ \textless{}number\textgreater{},\ "fat\_g":\ \textless{}number\textgreater{}\}}
\end{quote}

\textbf{(3a) Both photos, simple prompt} (exploratory):

\begin{quote}
\texttt{The\ first\ photo\ shows\ a\ meal\ before\ eating.\ The\ second\ photo\ shows\ the\ same\ tray\ after\ the\ person\ finished\ eating.}\strut \\
\texttt{Estimate\ the\ total\ food\ energy\ the\ person\ actually\ ate.}\strut \\
\texttt{Reply\ with\ JSON\ only:\ \{"kcal":\ \textless{}number\ eaten\textgreater{},\ "mass\_g":\ \textless{}grams\ of\ food\ eaten\textgreater{}\}}
\end{quote}

\textbf{(3b) Before photo only, structured prompt} (exploratory):

\begin{quote}
\texttt{This\ photo\ shows\ a\ meal\ before\ eating.\ Estimate\ the\ food\ energy\ that\ was\ served,\ the\ food\ energy\ that\ will\ be\ left\ over,}\strut \\
\texttt{and\ the\ food\ energy\ the\ person\ will\ actually\ eat\ (served\ minus\ left\ over).\ Also\ estimate\ the\ protein,\ carbohydrate\ and\ fat}\strut \\
\texttt{that\ the\ person\ will\ actually\ eat.}\strut \\
\texttt{Reply\ with\ JSON\ only:\ \{"served\_kcal":\ \textless{}number\textgreater{},\ "leftover\_kcal":\ \textless{}number\textgreater{},\ "kcal":\ \textless{}number\ eaten\textgreater{},\ "protein\_g":\ \textless{}number\textgreater{},\ "carbs\_g":\ \textless{}number\textgreater{},\ "fat\_g":\ \textless{}number\textgreater{}\}}
\end{quote}

Model API identifiers: claude-haiku-5-5, claude-sonnet-5-5, claude-opus-5-5 (Anthropic Messages API);
gpt-6-luna, gpt-6.1-sol, gpt-6-astra (OpenAI Chat Completions API); gemini-3.8-flash, gemini-4-argon (Google
Gemini API, google-genai SDK). Anthropic and OpenAI calls were made on 16 October 2026, Gemini calls on 20
October 2026.

Settings: low reasoning effort (Claude: adaptive thinking with effort ``low''; GPT: reasoning\_effort ``low'';
Gemini: thinking\_level ``low''), except the high-effort runs in Section 5.5 (effort or thinking\_level
``high''); temperature left at each vendor's default; images as PNG, long side ≤ 1,024 px; GPT images with
detail ``high''. Unparsable answers (malformed JSON) per system, before only / subtraction / both photos /
(3a) / (3b): Haiku 0/0/2/6/0, Sonnet 2/2/11/0/10, Opus 1/1/0/1/0, GPT-6 luna 0/0/0/0/0, GPT-6.1 sol 0/0/1/0/0,
GPT-6 astra 0/0/0/0/0, Gemini 3.8 Flash 0/0/0/0/0, Gemini 4 Argon 0/0/0/0/0. Per-call cost records for every
run are in the repository.

\section*{Appendix B. Pre-registration and deviations}\label{appendix-b.-pre-registration-and-deviations}
\addcontentsline{toc}{section}{Appendix B. Pre-registration and deviations}

The plan (questions Q1--Q3, prompts, thresholds, the 10\% leftover cut point, the pilot and its continue rule,
and the budget) was committed to the study repository with a timestamp before any dual-photo call. All steps
took place on 16 October 2026 (US Central time): pre-registration committed 13:07; pilot run 13:10; pilot
result and full-run candidates committed 13:11; subtraction condition committed 13:12; full run 13:13--14:05;
main results committed 14:09; the exploratory analyses in later commits (14:22 to 22:19); the ablation, repeat
and high-effort runs were registered at 22:19 and run 22:21--23:24, with results committed at 23:24. The
repository's commit history is the record.

\begin{itemize}
\tightlist
\item
  \textbf{Pilot} (100 meals drawn with a fixed seed; Haiku 5.5 and GPT-6.1 sol). Continue rule: for GPT-6.1
  sol, the dual-photo bias closer to zero than the before-only bias by at least 5 points, and Spearman
  correlation of predicted and true share eaten at least 0.3. Result: bias +12.3\% → −1.9\%, correlation 0.78;
  rule met. The prompt was not changed, so the pilot meals are kept.
\item
  \textbf{Deviation before the full run.} After a literature check, we added the subtraction condition
  (Section 4) as a secondary comparator. It was committed before any full-run call.
\item
  \textbf{Added after a second review (exploratory, registered before the calls):} conditions (3a) and (3b),
  the repeated calls and the high-effort runs.
\item
  \textbf{Third vendor, added after the first version of the paper (20 October 2026).} A selection rule, the
  same seven runs per model, the effort mapping and a budget cap were committed at 11:09 (US Central) before
  any Gemini call. Rule: the newest generally available, non-preview Flash model and the newest generally
  available, non-preview Pro model listed by the API on the run day. Flash resolved to Gemini 3.8 Flash. The
  Pro line had no generally available release newer than Gemini 2.5 Pro (June 2025; Gemini 3.1 Pro was preview
  only), while the vendor's pricing page listed Gemini 4 Argon as its flagship model; we took the flagship as
  the counterpart of Opus 5.5 and GPT-6 astra and recorded this as a deviation from the literal rule, decided
  before any Argon call. Runs took place 11:21--13:19 and all results were committed afterwards; no prompt,
  metric or threshold was changed.
\item
  \textbf{Added after the main results (exploratory):} direct test of the leftover-group difference;
  decomposition of the before-only error; model differences against Haiku 5.5; equivalence check of the day
  mean; local recalibration; leftover definition details and share-eaten errors; results by meal;
  Bland--Altman figure and the linear fit of day error on intake.
\end{itemize}

\section*{Appendix C. Full day-level results including
subtraction}\label{appendix-c.-full-day-level-results-including-subtraction}
\addcontentsline{toc}{section}{Appendix C. Full day-level results including subtraction}

Mean day error \% {[}95\% CI{]}; 95\% LoA \%; days within 800 kJ \%.

{\def\LTcaptype{none} 
\begin{longtable}[]{@{}
  >{\raggedright\arraybackslash}p{(\linewidth - 6\tabcolsep) * \real{0.2500}}
  >{\raggedright\arraybackslash}p{(\linewidth - 6\tabcolsep) * \real{0.2500}}
  >{\raggedright\arraybackslash}p{(\linewidth - 6\tabcolsep) * \real{0.2500}}
  >{\raggedright\arraybackslash}p{(\linewidth - 6\tabcolsep) * \real{0.2500}}@{}}
\toprule\noalign{}
\begin{minipage}[b]{\linewidth}\raggedright
Model
\end{minipage} & \begin{minipage}[b]{\linewidth}\raggedright
Before only
\end{minipage} & \begin{minipage}[b]{\linewidth}\raggedright
Subtraction
\end{minipage} & \begin{minipage}[b]{\linewidth}\raggedright
Both photos
\end{minipage} \\
\midrule\noalign{}
\endhead
\bottomrule\noalign{}
\endlastfoot
Haiku 5.5 & −4.2 {[}−8.9,~+0.7{]}; −48 to +39; 34 & −32.9 {[}−35.7,~−30.3{]}; −62 to −4; 8 & −12.0
{[}−16.2,~−7.2{]}; −53 to +29; 29 \\
Sonnet 5.5 & +5.5 {[}−0.7,~+12.3{]}; −62 to +73; 37 & −16.1 {[}−18.8,~−13.4{]}; −48 to +15; 28 & −0.6
{[}−4.2,~+3.2{]}; −41 to +40; 41 \\
Opus 5.5 & +25.4 {[}+19.4,~+32.3{]}; −32 to +83; 25 & +3.7 {[}+0.9,~+6.5{]}; −29 to +37; 49 & +12.0
{[}+8.7,~+15.5{]}; −25 to +49; 43 \\
GPT-6 luna & +6.7 {[}+1.9,~+11.8{]}; −41 to +54; 39 & −16.5 {[}−19.4,~−13.5{]}; −49 to +16; 22 & −4.3
{[}−7.4,~−1.0{]}; −38 to +30; 43 \\
GPT-6.1 sol & +17.7 {[}+12.6,~+23.4{]}; −33 to +69; 29 & −9.4 {[}−12.5,~−6.3{]}; −43 to +24; 30 & +3.4
{[}+0.3,~+6.5{]}; −31 to +38; 51 \\
GPT-6 astra & +20.2 {[}+15.0,~+25.9{]}; −31 to +71; 24 & −5.5 {[}−8.5,~−2.6{]}; −38 to +27; 42 & +7.6
{[}+4.4,~+10.8{]}; −27 to +42; 45 \\
Gemini 3.8 Flash & +10.6 {[}+5.8,~+16.0{]}; −36 to +57; 39 & −6.5 {[}−8.8,~−4.2{]}; −33 to +20; 49 & −0.5
{[}−2.9,~+2.0{]}; −28 to +27; 57 \\
Gemini 4 Argon & +20.4 {[}+14.9,~+26.4{]}; −32 to +72; 28 & +3.9 {[}+1.4,~+6.5{]}; −25 to +33; 45 & +4.5
{[}+1.9,~+7.1{]}; −25 to +34; 55 \\
\end{longtable}
}

\section*{Appendix D. Local recalibration and results by
meal}\label{appendix-d.-local-recalibration-and-results-by-meal}
\addcontentsline{toc}{section}{Appendix D. Local recalibration and results by meal}

\textbf{Local recalibration} (both photos; two constants refit on 20 meals from other participants; mean of
200 draws; metrics on the remaining participants).

{\def\LTcaptype{none} 
\begin{longtable}[]{@{}
  >{\raggedright\arraybackslash}p{(\linewidth - 8\tabcolsep) * \real{0.2000}}
  >{\raggedright\arraybackslash}p{(\linewidth - 8\tabcolsep) * \real{0.2000}}
  >{\raggedright\arraybackslash}p{(\linewidth - 8\tabcolsep) * \real{0.2000}}
  >{\raggedright\arraybackslash}p{(\linewidth - 8\tabcolsep) * \real{0.2000}}
  >{\raggedright\arraybackslash}p{(\linewidth - 8\tabcolsep) * \real{0.2000}}@{}}
\toprule\noalign{}
\begin{minipage}[b]{\linewidth}\raggedright
Model
\end{minipage} & \begin{minipage}[b]{\linewidth}\raggedright
Meal MAE\%: none → cal
\end{minipage} & \begin{minipage}[b]{\linewidth}\raggedright
Day mean \%: none → cal
\end{minipage} & \begin{minipage}[b]{\linewidth}\raggedright
Day 95\% LoA: none → cal
\end{minipage} & \begin{minipage}[b]{\linewidth}\raggedright
Days within 800 kJ \%: none → cal
\end{minipage} \\
\midrule\noalign{}
\endhead
\bottomrule\noalign{}
\endlastfoot
Haiku 5.5 & 24.9 → 25.0 & −12.0 → −4.5 & −53..+29 → −50..+41 & 29 → 32 \\
Sonnet 5.5 & 21.8 → 23.2 & −0.7 → −0.9 & −41..+40 → −42..+41 & 41 → 39 \\
Opus 5.5 & 20.5 → 19.1 & +12.1 → +0.5 & −25..+49 → −33..+34 & 43 → 43 \\
GPT-6 luna & 21.0 → 22.3 & −4.2 → −1.9 & −38..+29 → −37..+34 & 43 → 40 \\
GPT-6.1 sol & 19.5 → 20.8 & +3.4 → −1.1 & −31..+38 → −36..+33 & 51 → 41 \\
GPT-6 astra & 19.1 → 19.3 & +7.7 → −1.1 & −27..+43 → −34..+32 & 45 → 42 \\
Gemini 3.8 Flash & 16.6 → 17.8 & −0.5 → −1.0 & −27..+26 → −29..+27 & 57 → 52 \\
Gemini 4 Argon & 18.2 → 18.9 & +4.5 → +0.1 & −25..+34 → −29..+29 & 56 → 47 \\
\end{longtable}
}

\textbf{By meal} (both photos; MAE\% / bias \%).

{\def\LTcaptype{none} 
\begin{longtable}[]{@{}llll@{}}
\toprule\noalign{}
Model & Breakfast (n = 286) & Lunch (n = 257) & Dinner (n = 259) \\
\midrule\noalign{}
\endhead
\bottomrule\noalign{}
\endlastfoot
Haiku 5.5 & 19.0 / −6.7 & 22.4 / −6.2 & 31.4 / −25.8 \\
Sonnet 5.5 & 18.3 / +3.1 & 18.8 / −1.6 & 26.9 / −8.6 \\
Opus 5.5 & 20.6 / +12.1 & 20.0 / +12.3 & 20.8 / +4.1 \\
GPT-6 luna & 19.1 / −3.5 & 19.5 / +1.8 & 23.8 / −14.3 \\
GPT-6.1 sol & 19.8 / +5.0 & 18.2 / +10.0 & 20.4 / −8.7 \\
GPT-6 astra & 20.3 / +8.8 & 17.9 / +11.7 & 19.0 / −2.4 \\
Gemini 3.8 Flash & 15.9 / +1.6 & 15.2 / +5.3 & 18.4 / −11.2 \\
Gemini 4 Argon & 19.9 / +8.7 & 17.8 / +4.0 & 17.3 / −2.2 \\
\end{longtable}
}

Dinner, the largest meal, had the largest error for Haiku 5.5, Sonnet 5.5, GPT-6 luna and Gemini 3.8 Flash, in
line with the fall of the day error with rising intake.

\section*{Appendix E. Background experiment: optimizing on a public
benchmark}\label{appendix-e.-background-experiment-optimizing-on-a-public-benchmark}
\addcontentsline{toc}{section}{Appendix E. Background experiment: optimizing on a public benchmark}

Before this study we ran a separate experiment with a different target: energy \emph{served}, from the before
photo only. We used an agentic search (LLM coding agents proposing and scoring prompt-and-code pipelines) over
three generations on Nutrition5k overhead photos {[}17{]} with Claude Haiku 5.5. On the held-out Nutrition5k
test split (378 dishes) the searched pipeline reduced MAE\% from 42.4\% (direct prompt) to 27.1\%, but a
direct prompt plus a two-constant calibration fitted on 300 training dishes came within 1.0--2.4 points of it
for six models. On the before photos of ACETADA, with served energy as the label, the same benchmark-fitted
calibration made every system worse (for Haiku 5.5, direct prompt 24.8\% vs searched pipeline 40.0\%). Because
the target and the input differ from the present study, we do not use these results for its conclusions; full
details are in the repository.

\section*{Appendix F. Fixed-fraction baseline and sensitivity
analyses}\label{appendix-f.-fixed-fraction-baseline-and-sensitivity-analyses}
\addcontentsline{toc}{section}{Appendix F. Fixed-fraction baseline and sensitivity analyses}

\textbf{Fixed eaten fraction} (before-only estimate × c, c from other participants; MAE\% and day-level
results).

{\def\LTcaptype{none} 
\begin{longtable}[]{@{}
  >{\raggedright\arraybackslash}p{(\linewidth - 12\tabcolsep) * \real{0.1429}}
  >{\raggedright\arraybackslash}p{(\linewidth - 12\tabcolsep) * \real{0.1429}}
  >{\raggedright\arraybackslash}p{(\linewidth - 12\tabcolsep) * \real{0.1429}}
  >{\raggedright\arraybackslash}p{(\linewidth - 12\tabcolsep) * \real{0.1429}}
  >{\raggedright\arraybackslash}p{(\linewidth - 12\tabcolsep) * \real{0.1429}}
  >{\raggedright\arraybackslash}p{(\linewidth - 12\tabcolsep) * \real{0.1429}}
  >{\raggedright\arraybackslash}p{(\linewidth - 12\tabcolsep) * \real{0.1429}}@{}}
\toprule\noalign{}
\begin{minipage}[b]{\linewidth}\raggedright
Model
\end{minipage} & \begin{minipage}[b]{\linewidth}\raggedright
MAE\% before×c / both
\end{minipage} & \begin{minipage}[b]{\linewidth}\raggedright
Δ both − before×c, all {[}95\% CI{]}
\end{minipage} & \begin{minipage}[b]{\linewidth}\raggedright
Δ, ≥10\% left {[}95\% CI{]}
\end{minipage} & \begin{minipage}[b]{\linewidth}\raggedright
Day mean \%: before×c / both
\end{minipage} & \begin{minipage}[b]{\linewidth}\raggedright
Day LoA \%: before×c / both
\end{minipage} & \begin{minipage}[b]{\linewidth}\raggedright
Days within 800 kJ \%: before×c / both
\end{minipage} \\
\midrule\noalign{}
\endhead
\bottomrule\noalign{}
\endlastfoot
Haiku 5.5 & 27.3 / 24.9 & −2.5 {[}−4.2,~−0.6{]} & −1.8 {[}−4.4,~+0.6{]} & −16.6 / −12.0 & −55..+22 / −53..+29
& 28 / 29 \\
Sonnet 5.5 & 24.9 / 21.8 & −3.0 {[}−5.5,~−0.8{]} & −1.3 {[}−5.9,~+2.7{]} & −8.2 / −0.6 & −67..+51 / −41..+40 &
32 / 41 \\
Opus 5.5 & 22.7 / 20.5 & −2.2 {[}−4.8,~+0.3{]} & −7.5 {[}−12.3,~−3.3{]} & +9.2 / +12.0 & −41..+60 / −25..+49 &
35 / 43 \\
GPT-6 luna & 23.7 / 21.0 & −2.6 {[}−4.2,~−1.1{]} & −1.3 {[}−5.1,~+1.9{]} & −7.1 / −4.3 & −48..+34 / −38..+30 &
27 / 43 \\
GPT-6.1 sol & 22.3 / 19.5 & −2.8 {[}−4.8,~−0.7{]} & −6.6 {[}−10.6,~−3.1{]} & +2.5 / +3.4 & −42..+47 / −31..+38
& 41 / 51 \\
GPT-6 astra & 22.4 / 19.0 & −3.4 {[}−5.3,~−1.5{]} & −8.8 {[}−12.7,~−5.1{]} & +4.7 / +7.6 & −40..+49 / −27..+42
& 40 / 45 \\
Gemini 3.8 Flash & 20.4 / 16.7 & −3.7 {[}−5.2,~−2.3{]} & −5.4 {[}−8.4,~−2.8{]} & −3.7 / −0.5 & −45..+37 /
−28..+27 & 41 / 57 \\
Gemini 4 Argon & 22.6 / 18.2 & −4.4 {[}−6.2,~−2.6{]} & −7.7 {[}−11.2,~−4.4{]} & +4.8 / +4.5 & −41..+50 /
−25..+34 & 39 / 55 \\
\end{longtable}
}

\textbf{Leftover effect modification on the kcal scale}, d = \textbar both − eaten\textbar{} − \textbar before
− eaten\textbar, kcal per meal {[}95\% CI{]}.

{\def\LTcaptype{none} 
\begin{longtable}[]{@{}llll@{}}
\toprule\noalign{}
Model & ≥10\% left & \textless10\% left & Difference \\
\midrule\noalign{}
\endhead
\bottomrule\noalign{}
\endlastfoot
Haiku 5.5 & −28 {[}−47,~−8{]} & +20 {[}+4,~+40{]} & −48 {[}−75,~−22{]} \\
Sonnet 5.5 & −43 {[}−78,~−9{]} & +2 {[}−21,~+21{]} & −45 {[}−87,~−3{]} \\
Opus 5.5 & −145 {[}−181,~−110{]} & −8 {[}−29,~+8{]} & −138 {[}−178,~−97{]} \\
GPT-6 luna & −45 {[}−76,~−14{]} & −1 {[}−10,~+7{]} & −43 {[}−74,~−12{]} \\
GPT-6.1 sol & −116 {[}−147,~−84{]} & −3 {[}−16,~+13{]} & −113 {[}−146,~−79{]} \\
GPT-6 astra & −135 {[}−166,~−102{]} & −8 {[}−16,~−0{]} & −127 {[}−157,~−93{]} \\
Gemini 3.8 Flash & −90 {[}−116,~−63{]} & +2 {[}−5,~+9{]} & −92 {[}−119,~−64{]} \\
Gemini 4 Argon & −134 {[}−163,~−102{]} & −20 {[}−31,~−9{]} & −114 {[}−146,~−82{]} \\
\end{longtable}
}

\textbf{Without the pilot.} Excluding the 100 pilot meals leaves 702 meals and 100 complete days; excluding
all pilot participants leaves 371 meals and 64 complete days. MAE\% change (both − before), all meals: Haiku
5.5 +0.3 / −0.8, Sonnet 5.5 −3.1 / −2.7, Opus 5.5 −10.0 / −10.4, GPT-6 luna −2.3 / −3.3, GPT-6.1 sol −7.3 /
−7.9, GPT-6 astra −8.7 / −9.3, Gemini 3.8 Flash −5.0 / −4.7, Gemini 4 Argon −9.6 / −10.5.

\textbf{Between-model differences in Δ MAE\% (both − before), all 28 pairs, unadjusted {[}95\% CI{]}.}
Haiku−Sonnet +2.5 {[}−0.1,~+5.2{]}; Haiku−Opus +9.7 {[}+6.7,~+12.9{]}; Haiku−luna +3.0 {[}+1.0,~+5.1{]};
Haiku−sol +7.5 {[}+4.9,~+10.3{]}; Haiku−astra +8.7 {[}+6.5,~+11.2{]}; Haiku−Flash +5.4 {[}+3.2,~+7.7{]};
Haiku−Argon +9.8 {[}+7.4,~+12.5{]}; Sonnet−Opus +7.2 {[}+3.7,~+10.7{]}; Sonnet−luna +0.4 {[}−2.3,~+2.9{]};
Sonnet−sol +5.0 {[}+2.3,~+7.5{]}; Sonnet−astra +6.2 {[}+3.5,~+8.6{]}; Sonnet−Flash +2.8 {[}+0.2,~+5.3{]};
Sonnet−Argon +7.2 {[}+4.5,~+9.9{]}; Opus−luna −6.8 {[}−9.5,~−4.2{]}; Opus−sol −2.2 {[}−4.8,~−0.0{]};
Opus−astra −1.0 {[}−3.3,~+1.0{]}; Opus−Flash −4.4 {[}−6.7,~−2.4{]}; Opus−Argon +0.1 {[}−2.4,~+2.3{]}; luna−sol
+4.5 {[}+2.1,~+6.9{]}; luna−astra +5.8 {[}+3.9,~+7.8{]}; luna−Flash +2.4 {[}+0.7,~+4.2{]}; luna−Argon +6.8
{[}+5.0,~+8.8{]}; sol−astra +1.2 {[}−0.0,~+2.7{]}; sol−Flash −2.1 {[}−3.7,~−0.4{]}; sol−Argon +2.3
{[}+0.2,~+4.7{]}; astra−Flash −3.4 {[}−4.7,~−2.2{]}; astra−Argon +1.1 {[}−0.6,~+2.7{]}; Flash−Argon +4.4
{[}+2.9,~+6.0{]}.

\textbf{Day-level agreement, repeated-measures and log-scale} (both photos). Within-participant SD of e\_day /
ICC / ratio LoA (estimate ÷ truth): Haiku 5.5 16.3 / 0.41 / ×0.57--×1.30; Sonnet 5.5 18.3 / 0.22 /
×0.67--×1.42; Opus 5.5 12.9 / 0.53 / ×0.81--×1.51; GPT-6 luna 13.3 / 0.40 / ×0.67--×1.33; GPT-6.1 sol 13.4 /
0.43 / ×0.74--×1.41; GPT-6 astra 12.7 / 0.49 / ×0.78--×1.46; Gemini 3.8 Flash 11.0 / 0.37 / ×0.76--×1.29;
Gemini 4 Argon 13.8 / 0.15 / ×0.78--×1.37. ICC range 0.15 to 0.53.

\textbf{Protein error and meal composition} (both photos; Spearman with protein density g/100 kcal, Spearman
with animal-source energy share; bias in meals with animal share \textless1/3 vs ≥1/3): Haiku 5.5 −0.50,
−0.35; −20\% vs −37\%. Sonnet 5.5 −0.50, −0.31; −16\% vs −31\%. Opus 5.5 −0.46, −0.27; −2\% vs −15\%. GPT-6
luna −0.47, −0.24; −13\% vs −24\%. GPT-6.1 sol −0.41, −0.11; −12\% vs −20\%. GPT-6 astra −0.38, −0.10; −9\% vs
−15\%. Gemini 3.8 Flash −0.42, −0.20; −15\% vs −26\%. Gemini 4 Argon −0.38, −0.09; −2\% vs −8\%.

\textbf{Recall methods, approximate error range.} From the trial's mean (SD) of estimated ÷ true energy in \%,
(mean − 100) ± 1.96 SD: ASA24 −52 to +62; Intake24 −55 to +59; mobile food record with trained analyst −28 to
+31; image-assisted interviewer recall −24 to +54. These are derived from rounded summary statistics, not from
raw data, and are not compared with our limits.

\textbf{Day level for the ablation conditions} (mean e\_day \%; 95\% LoA; days within 800 kJ \%). (1) / (3b) /
(3a) / (3): Haiku 5.5 −4.2, −48..+39, 34 / −15.5, −49..+18, 28 / −12.5, −54..+29, 32 / −12.0, −53..+29, 29.
Sonnet 5.5 +5.5, −62..+73, 37 / −0.4, −63..+62, 40 / −5.7, −36..+24, 41 / −0.6, −41..+40, 41. Opus 5.5 +25.4,
−32..+83, 25 / +12.5, −33..+58, 33 / +8.0, −25..+41, 50 / +12.0, −25..+49, 43. GPT-6 luna +6.7, −41..+54, 39 /
+6.0, −40..+52, 32 / −5.7, −39..+27, 39 / −4.3, −38..+30, 43. GPT-6.1 sol +17.7, −33..+69, 29 / +20.0,
−31..+71, 28 / −0.5, −31..+30, 48 / +3.4, −31..+38, 51. GPT-6 astra +20.2, −31..+71, 24 / +22.5, −27..+72, 25
/ +3.7, −29..+36, 47 / +7.6, −27..+42, 45. Gemini 3.8 Flash +10.6, −36..+57, 39 / +12.5, −34..+59, 37 / −2.1,
−28..+24, 53 / −0.5, −28..+27, 57. Gemini 4 Argon +20.4, −32..+72, 28 / −3.5, −42..+35, 36 / −0.5, −31..+30,
47 / +4.5, −25..+34, 55.

\textbf{Grams eaten from (3a)} (MAE g; error relative to mean true grams; bias): Haiku 5.5 274 g, 37.9\%,
−36.7\%; Sonnet 5.5 185 g, 25.6\%, −21.9\%; Opus 5.5 149 g, 20.6\%, −11.9\%; GPT-6 luna 268 g, 37.1\%,
−36.0\%; GPT-6.1 sol 183 g, 25.2\%, −21.9\%; GPT-6 astra 171 g, 23.6\%, −19.5\%; Gemini 3.8 Flash 166 g,
23.0\%, −20.4\%; Gemini 4 Argon 135 g, 18.6\%, −12.5\%.

\textbf{Test--retest, condition (3), 100 pilot meals, three calls} (MAE\% per call; median per-meal CV; mean
absolute difference between two calls as \% of truth): Haiku 5.5 29.3 / 28.8 / 28.2; 8.2\%; 11.4\%. Sonnet 5.5
21.4 / 21.4 / 20.1; 4.6\%; 8.2\% (96 meals parsed in all three). Opus 5.5 23.3 / 21.9 / 23.1; 5.9\%; 8.7\%.
GPT-6 luna 23.7 / 22.4 / 23.5; 7.6\%; 11.9\% (99 meals parsed in all three). GPT-6.1 sol 20.5 / 20.7 / 21.2;
3.5\%; 6.6\%. GPT-6 astra 20.5 / 21.0 / 21.1; 3.8\%; 6.1\%. Gemini 3.8 Flash 18.9 / 19.8 / 19.4; 4.7\%; 7.5\%.
Gemini 4 Argon 18.8 / 20.2 / 19.4; 4.2\%; 8.5\%.

\textbf{Reasoning effort, condition (3), same 100 meals} (MAE\% low, mean of three calls → high; bias low →
high; cost per meal, high relative to low): Haiku 5.5 28.8 → 24.6; −15.7 → −12.6; ×1.7. Sonnet 5.5 21.0 →
20.7; −7.3 → −9.3; ×1.1. Opus 5.5 22.7 → 22.3; +7.5 → +4.7; ×1.2. GPT-6 luna 23.2 → 24.1; −7.6 → −8.4; ×2.2.
GPT-6.1 sol 20.8 → 21.1; −1.9 → −0.3; ×2.5. GPT-6 astra 20.9 → 21.4; +3.1 → +3.1; ×2.6. Gemini 3.8 Flash 19.4
→ 19.2; −6.3 → −6.7; ×3.4. Gemini 4 Argon 19.5 → 19.4; −0.7 → +0.7; ×4.3.

\textbf{Valid-output subset} (meals parsed in all three conditions for that model): MAE\% before / subtraction
/ both --- Haiku 5.5 (n = 800) 24.7 / 36.5 / 24.1; Sonnet 5.5 (n = 789) 23.1 / 24.9 / 19.3; Opus 5.5 (n = 801)
29.4 / 21.0 / 20.5; GPT-6 luna (n = 802) 23.9 / 26.2 / 21.0; GPT-6.1 sol (n = 801) 26.9 / 23.8 / 19.1; GPT-6
astra (n = 802) 27.6 / 22.5 / 19.0; Gemini 3.8 Flash (n = 802) 21.9 / 18.5 / 16.7; Gemini 4 Argon (n = 802)
27.9 / 21.2 / 18.2.

\section*{References}\label{references}
\addcontentsline{toc}{section}{References}

{[}1{]} Subar AF, Freedman LS, Tooze JA, et al.~Addressing Current Criticism Regarding the Value of
Self-Report Dietary Data. The Journal of Nutrition 145(12):2639--2645. 2015. doi:10.3945/jn.115.219634.

{[}2{]} Boushey CJ, Spoden M, Zhu FM, Delp EJ, Kerr DA. New mobile methods for dietary assessment: review of
image-assisted and image-based dietary assessment methods. Proceedings of the Nutrition Society
76(3):283--294. 2017. doi:10.1017/S0029665116002913.

{[}3{]} Whitton C, Collins CE, Mullan BA, et al.~Accuracy of energy and nutrient intake estimation versus
observed intake using 4 technology-assisted dietary assessment methods: a randomized crossover feeding study.
The American Journal of Clinical Nutrition 120(1):196--210. 2024. doi:10.1016/j.ajcnut.2024.04.030.

{[}4{]} Whitton C, Healy JD, Collins CE, et al.~Accuracy and Cost-effectiveness of Technology-Assisted Dietary
Assessment Comparing the Automated Self-administered Dietary Assessment Tool, Intake24, and an Image-Assisted
Mobile Food Record 24-Hour Recall Relative to Observed Intake: Protocol for a Randomized Crossover Feeding
Study. JMIR Research Protocols 10(12):e32891. 2021. doi:10.2196/32891.

{[}5{]} Coburn B, He J, Rollo ME, et al.~Evaluating Large Multimodal Models for Nutrition Analysis: A New
Benchmark Enriched with Contextual Metadata. IEEE EMBS International Conference on Biomedical and Health
Informatics (BHI), pp.~1--7. 2025. doi:10.1109/BHI67747.2025.11269555. Extended version: arXiv:2507.07048.
Dataset: https://skynet.ecn.purdue.edu/\textasciitilde coburn6/ACETADA/ (CC BY-NC 4.0).

{[}6{]} Vinod G, Raghavan S, Coburn B, Zhu F. DietDelta: A Vision-Language Approach for Dietary Assessment via
Before-and-After Images. Proceedings of the IEEE/CVF Conference on Computer Vision and Pattern Recognition
(CVPR) Workshops, pp.~9591--9600. 2026. arXiv:2604.06352.

{[}7{]} Nakagawa S, Yamamoto A. Effects of Dish Name, Visible Amount, Persona, and Explicit Instruction on
Energy Estimates From Food Photographs by 26 Vision-Language Models: Comparative Evaluation Study. medRxiv
preprint (not peer reviewed). 2026. doi:10.64898/2026.10.09.26365134.

{[}8{]} Lo FP-W, Qiu J, Wang Z, et al.~Dietary Assessment With Multimodal ChatGPT: A Systematic Analysis. IEEE
Journal of Biomedical and Health Informatics 28(12):7577--7587. 2024. doi:10.1109/JBHI.2024.3417280.

{[}9{]} Chen YJ, Chang C-C, Hoang YN, et al.~Customized multimodal Diabot-GPT-4o enhances accuracy of
image-based dietary assessments in dietetic trainees in Taiwan: validation against weighed food records. The
American Journal of Clinical Nutrition 122(6):1836--1849. 2025. doi:10.1016/j.ajcnut.2025.10.013.

{[}10{]} Lu Y, Stathopoulou T, Vasiloglou MF, et al.~An Artificial Intelligence-Based System to Assess
Nutrient Intake for Hospitalised Patients. IEEE Transactions on Multimedia 23:1136--1147. 2021.
doi:10.1109/TMM.2020.2993948.

{[}11{]} Larke JA, Chin EL, Bouzid YY, et al.~Surveying Nutrient Assessment with Photographs of Meals
(SNAPMe): A Benchmark Dataset of Food Photos for Dietary Assessment. Nutrients 15(23):4972. 2023.
doi:10.3390/nu15234972.

{[}12{]} Sterling S, et al.~Vision-Language Models for Image-Based Dietary Assessment: A Benchmark of
Accuracy, Cost, and Prompt Strategies Across Ten Models. bioRxiv preprint. 2026.
doi:10.64898/2026.07.26.740845.

{[}13{]} Fridolfsson J, et al.~Performance Evaluation of 3 Large Language Models for Nutritional Content
Estimation from Food Images. Current Developments in Nutrition 9(10):107556. 2025.
doi:10.1016/j.cdnut.2025.107556.

{[}14{]} Coburn B, et al.~Open-KNEAD: Knowledge-grounded Nutrition Estimation via Agentic Decomposition. arXiv
preprint. 2026. arXiv:2607.12911.

{[}15{]} Bland JM, Altman DG. Statistical methods for assessing agreement between two methods of clinical
measurement. Lancet 1(8476):307--310. 1986. doi:10.1016/S0140-6736(86)90837-8.

{[}16{]} Schuirmann DJ. A comparison of the two one-sided tests procedure and the power approach for assessing
the equivalence of average bioavailability. Journal of Pharmacokinetics and Biopharmaceutics 15(6):657--680.
1987. doi:10.1007/BF01068419.

{[}17{]} Thames Q, Karpur A, Norris W, et al.~Nutrition5k: Towards Automatic Nutritional Understanding of
Generic Food. CVPR, pp.~8899--8907. 2021. doi:10.1109/CVPR46437.2021.00879.

{[}18{]} Van Calster B, McLernon DJ, van Smeden M, et al.~Calibration: the Achilles heel of predictive
analytics. BMC Medicine 17:230. 2019. doi:10.1186/s12916-019-1466-7.

{[}19{]} Janssen KJM, Moons KGM, Kalkman CJ, Grobbee DE, Vergouwe Y. Updating methods improved the performance
of a clinical prediction model in new patients. Journal of Clinical Epidemiology 61(1):76--86. 2008.
doi:10.1016/j.jclinepi.2007.04.018.

{[}20{]} Piccoli F, Callegaro D, Marelli D, et al.~FLIC: A Real-World Dataset for Visual Estimation of Food
Leftovers in Canteens. Applied Sciences 16(11):5465. 2026. doi:10.3390/app16115465.

{[}21{]} Bland JM, Altman DG. Agreement between methods of measurement with multiple observations per
individual. Journal of Biopharmaceutical Statistics 17(4):571--582. 2007. doi:10.1080/10543400701329422.

{[}22{]} Krouwer JS. Why Bland--Altman plots should use X, not (Y+X)/2 when X is a reference method.
Statistics in Medicine 27(5):778--780. 2008. doi:10.1002/sim.3086.

\end{document}